\documentclass[twocolumn]{webofc}

\usepackage[varg]{txfonts}   
\newcommand{\avg}[1]{\left\langle#1\right\rangle}

\begin{document}
\title{Kinetic roughening in active interfaces}
%
%

\author{\firstname{Francesco} \lastname{Cagnetta}\inst{1}\fnsep\thanks{\email{F.Cagnetta@ed.ac.uk}} \and
        \firstname{Martin R.} \lastname{Evans}\inst{1} \and
        \firstname{Davide} \lastname{Marenduzzo}\inst{1}
}

\institute{SUPA, School of Physics and Astronomy, University of Edinburgh, Peter Guthrie Tait Road, Edinburgh EH9 3FD, United Kingdom}

\abstract{%
  The essential features of many interfaces driven out of equilibrium are described by the same equation---the Kardar-Parisi-Zhang (KPZ) equation. How do living interfaces, such as the cell membrane, fit into this picture? In an endeavour to answer such a question, we proposed in [F. Cagnetta, M. R. Evans, D. Marenduzzo, PRL 120, 258001 (2018)] an idealised model for the membrane of a moving cell. Here we discuss how the addition of simple ingredients inspired by the dynamics of the membrane of moving cells affects common kinetic roughening theories such as the KPZ and Edwards-Wilkinson equations.
}
\maketitle
\section{Introduction}
\label{intro}

The theory of kinetic roughening describes the properties of surfaces which might appear smooth or rough depending on the scale of observation~\cite{halpin1995kinetic,barabasi1995}. Although usually applied to inanimate surfaces, such as solid-liquid interfaces, the theory can also describe biological, animate interfaces and this is the subject of the present paper. The Eden model, for instance, which represents one of the earliest attempt towards a probabilistic formulation of cluster growth, was introduced as an oversimplified portrayal of an expanding bacterial colony~\cite{eden1961growth}.

However, while some biological interfaces are simply understood in terms of scaling concepts, the features of some others have proven far more challenging to fathom. For instance, the issue of whether the low-frequency fluctuations in the shape of red blood cells are thermal or not has been resolved only recently~\cite{turlier2016aa}, after a forty year debate \cite{brochard1975frequency}.
It was originally believed, in fact, that the observed $1/k^4$ spectrum of fluctuations could be described by the energy equipartition principle~\cite{brochard1975frequency}.
However, Prost and Bruinsma~\cite{prost1996active}  showed that active fluctuations, generated by ATP-consuming processes, contribute to the dynamics of biological interfaces as well as thermal fluctuations. This initiated the field of active interfaces~\cite{ramaswamy2000aa,gov2006aa,maitra2014aa}: whithin such framework, the characteristics of fluctuations spectra (such as the $1/k^4$ behaviour in flickering) can be attributed to various membrane activities (such as that of ion channels). It was indeed shown, in~\cite{turlier2016aa}, that membrane response and fluctuations violated the fluctuation–dissipation relation, indicating the non-equilibrium nature of fluctuations. All this indicates that an understanding of active process is crucial for any quantitative description of membrane dynamics.

In the present work we are concerned with active processes that are related to cell locomotion. The perspective we take~\cite{cagnetta2018aa,cagnetta2019aa}, as in usual kinetic roughening theories, is focused on the scaling of the interface width $w$ with the system size~\cite{halpin1995kinetic}. We present a brief review of such theories (Edwards-Wilkinson and Kardar-Parisi-Zhang equations) in Section \ref{sec-0}.
In Section~\ref{sec-1} we introduce our modelling approach, inspired by the leading edge of eukaryotic cells crawling on two-dimensional substrates: the protruding force which sets the membrane in motion is exerted by the actin cytoskeleton below the membrane, but directed by membrane proteins which collect signals from the exterior of the cell.
The first model we discuss consists indeed of a moving interface whose growth is effected by a number of diffusing particles, which we refer to as \emph{activators}. The model is a linear theory for active interfaces which can be thought of as an extension of the Edwards-Wilkinson equation. In Section 4 we show that lattice simulations of growth with diffusing activators presents different scaling behaviour, which we ascribe to a non-linear effect analogous to that considered in the Kardar-Parisi-Zhang equation. Finally, in Section 5, we consider in detail the coupling between the interface shape and the activators distribution generated by active growth.

\section{Theory of kinetic roughening}\label{sec-0}
Consider an interface described by a time-dependent height function $h(x,t)$ over some substrate $x$. The average of $h(x,t)$ over $x$ yields the mean distance from the substrate, while the variance gives the squared width $w^2(L,t)$ about the mean profile, function of the system linear size $L$.
The width is related to the time-dependent structure factor (square modulus of the height Fourier modes $h_k(t)$) by
\begin{equation}\label{eq:widthdef}
w^2(L,t) = \frac{1}{L^{2d}}\sum_{k>0} \avg{\left|h_k(t)\right|^2}.
\end{equation}
In a system of linear size $L$, the smallest wavenumber allowed is $2\pi/L$: hence, the small-$k$ scaling of the structure factor translates into the scaling of the width with $L$, a fact often referred to through the statement that the interface roughness depends on the scale of observation~\cite{halpin1995kinetic}. In fact, not only the width depends on system size, but also the time it takes for the width to develop, starting from a flat interface with $w=0$. Both these facts are summarised in the Family-Vicsek scaling hypothesis~\cite{family1985scaling},
\begin{equation}\label{eq:familyvicsek}
w(L,t) = L^{\alpha}f(t/L^{z}),
\end{equation}
with roughness exponent $\alpha$, dynamic exponent $z$ and scaling function $f(x)$ which is constant at large $x$ and behaves as $x^{\alpha/z}$ for small $x$.

The simplest model satisfying the Family-Vicsek scaling is the Edwards-Wilkinson (EW) equation~\cite{edwards1982surface},
\begin{equation}\label{eq:edwardswilkinson}
\partial_t h(x,t) = \nu \nabla^2 h(x,t) + \sqrt{2D_h}\eta(x,t),
\end{equation}
with $\eta$ a gaussian, white, space-time noise with zero mean and unit variance. $\nu$ is the surface tension of the interface, $D_h$ quantifies the intensity of fluctuations.
Upon identifying $D_h$ with the temperature $k_BT$, the right hand side of (\ref{eq:edwardswilkinson}) represents the
competition between the smoothing effect of surface tension and random height fluctuations generated by thermal  noise.
By solving Eq.~(\ref{eq:edwardswilkinson}) for the modes of $h(x,t)$ in one spatial dimension, with flat initial condition $h(x,t=0)=0$, a simple calculation (see e.g.  \cite{barabasi1995}) shows that Eq.~(\ref{eq:familyvicsek}) is obeyed with $\alpha=1/2$ and $z=2$. Additionally, $z=2$ holds for all dimensions, while $\alpha=0$ in $d\geq 2$, implying a smooth interface.

When the interface is driven out of equilibrium, (\ref{eq:edwardswilkinson}) is augmented with additional terms. On the one hand, a uniform driving force, in the form of a constant term $\lambda$ added to the right-hand side of Eq.~(\ref{eq:edwardswilkinson}), can be removed with a galilean shift of the height $h\to h-\lambda t$, thus it would not alter the EW scaling. On the other hand, geometric considerations \cite{kardar1986aa} imply that driving forces are generically directed along the local normal to the interface, so that the $\lambda$-term aquires a projection factor $\sqrt{1+(\nabla h)^2}$ (cf. Fig.~\ref{fig-2}). Expanding the square root for $\nabla h$ small yields the celebrated Kardar-Parisi-Zhang (KPZ) equation~\cite{kardar1986aa}
\begin{equation}\label{eq:KPZ}
\partial_t h(x,t) = \nu \nabla^2 h(x,t) + \frac{\lambda}{2} (\nabla h)^2+ \sqrt{2D_h}\eta(x,t),
\end{equation}
 characterised by scaling exponents $\alpha=1/2$ and $z=3/2$ in one spatial dimension.

\section{Growth by diffusing activators: modified  EW equation}
\label{sec-1}

On the front of a moving cell, or leading edge, the driving force is localised around the positions of specific membrane proteins~\cite{prost1996active} which catalyse membrane growth~\cite{gov2006aa}.
As they are the source of active growth, we call these proteins \emph{activators} and denote their position with $X_i(t)$, $i=1,\dots,N$. 
We begin by assuming that \emph{i)} the plasma membrane can be described by an height function $h(x,t)$; \emph{ii)} the height fluctuations are described by (\ref{eq:edwardswilkinson}) in the absence of activators; \emph{iii)} the activators' motion within the interface is purely diffusive. We then write the following field equation for our active interface (and activators),
\begin{subequations}
\begin{equation}
\label{eq:activeedwardswilkinson}
\partial_t h(x,t) = \lambda \delta\left(x-X_i(t)\right) + \nu \nabla^2 h(x,t) 
+ \sqrt{2D_h}\eta(x,t)
\end{equation}
\begin{equation}
\label{eq:X}
\dot{X}_i(t) = \xi_i(t),
\end{equation}
\end{subequations}
with $\xi_i$ an independent Gaussian white noise with zero mean and variance $2D_a$ for each $i$. We stress that the diffusion coefficient of the activators $D_a$ is not related to the coefficient of the interface noise $\eta$, as the activator noise regards fluctuations of the positions $X_i$ within the interface, while the interface noise refers to fluctuations of the $d$-dimensional interface in the $d+1$ dimensional space.

Eq.~(\ref{eq:activeedwardswilkinson}) can be solved via Fourier transform $h_k(t)=\int_{[0,L]^d} d^dx\, h(x,t)e^{-ikx}$. With $N=1$ activator, the transform of Eq.~(\ref{eq:activeedwardswilkinson}) reads
\begin{equation}\label{eq:activeEWfourierD}
\dot{h}_k(t) = -\nu k^2 h_k(t) + \lambda e^{-ikX(t)} + \sqrt{2D_h}\eta_k(t),
\end{equation}
with $\avg{\eta_k(t)\eta_{k'}(t')} = L^d\delta\left(t-t'\right)\delta_{k,k'}$. By substituting $X(t) = X_0 + \int_0^t ds\, \xi(s)$ and integrating from a flat interface at $t=0$, we get
\begin{equation}\label{eq:activeEWfourierI}
 h_k(t) = \int_0^t ds\, e^{-\nu k^2(t- s)}\left[\eta_k(s)+\lambda e^{-ikX_0}e^{-ik\int_0^s du\, \xi(u)}\right].
\end{equation}
Upon averaging w.r.t the noise distribution, the $\eta_k$ term vanishes, while the average of the complex exponential coincides with the characteristic function of the variable $\int_0^s du\,\xi(u)$. As the latter variable is Gaussian with average zero and variance $2Ds$, the characteristic function is $e^{-2D_a s k^2}$. Thus, after inverting the Fourier transform, the average height profile reads
\begin{equation}\label{eq:activeEWaverage}
 \avg{h(x,t)} =\frac{\lambda}{L^d} \sum_{k} \frac{1-e^{-(\nu-D_a)k^2t}}{(\nu-D_a)k^2} e^{-D_ak^2t}e^{ik\left(x-X_0\right)},
\end{equation}
at variance with the vanishing average height of the EW equation. Eq.~(\ref{eq:activeEWaverage}) represents the convolution of the EW response function $(1-e^{-\nu k^2 t})/\nu k^2$, with $\nu$ replaced by $(\nu-D_a)$, with the probability density function $p_a(x,t)$ of the position $X(t)$ of the activator. Contributions due to different activators add up linearly.

Eq.~(\ref{eq:activeEWfourierI}) gives us access to the structure factor too, which reads,
\begin{equation}
\begin{aligned}
 \frac{1}{L^d}\avg{|h_k(t)|^2} = \frac{1}{L^d}\int_0^t ds_1\,\int_0^t ds_2\, e^{-\nu k^2(2t-s_1-s_2)} \times \\
 \left[ \avg{\eta_k(s_1)\eta_{-k}(s_2)} + \lambda^2 \avg{e^{-ik\int_{s_2}^{s_1} du\, \xi_i(u)}}\right].
\end{aligned}
\end{equation}
The second contribution in the square brackets coincides with the characteristic function of the Gaussian variable $\int_{s_2}^{s_1} du\, \xi(u)$, whose variance equals $2D|s_1-s_2|$. By plugging in also the $\eta$ noise correlations in the $k$-space, we get
\begin{equation}\label{eq:activeEWwidth}
 \begin{aligned}
  \frac{1}{L^d}\avg{|h_k(t)|^2} = \frac{D_h}{\nu k^2}\left(1-e^{-\nu k^2 t}\right) +\\
  \frac{\lambda^2}{L^d}\frac{\left[(\nu-D_a)+(\nu+D_a)e^{-2\nu k^2 t}-2\nu e^{-2(\nu+D_a) k^2 t}\right]}{\nu\left(\nu^2-D_a^2\right) k^4}.
 \end{aligned}
\end{equation}
The generalisation to multiple activators is simple, at least for the modes with $k\neq 0$ (the only modes relevant for the width, according to Eq.~(\ref{eq:widthdef})). Specifically, it suffices to multiply the second term
in Eq.~(\ref{eq:activeEWwidth}) by the number of activators $N$, so that the overall density $\rho_0 = N/L^d$ appears.

As time is always multiplied by $k^2$ in Eq.~(\ref{eq:activeEWaverage}), we conclude that $z=2$. To compute $\alpha$, we consider the $t\rightarrow\infty$ limit, where all the exponential factors vanish and we are left with
\begin{equation}\label{eq:activeEWstationarySF}
 \frac{1}{L^d}\avg{|h_k(t)|} \to \frac{D_h}{\nu k^2} +  \frac{\rho_0 \lambda^2}{\nu\left(\nu+D_a\right) k^4}.
\end{equation}
The small $k$ behaviour of Eq.~(\ref{eq:activeEWstationarySF}) is dominated by the $1/k^4$ term, so that, in the $L\rightarrow\infty$ limit, $w^2(L) \sim L^{4-d}$, i.e. $\alpha=(4-d)/2$ for $d<4$, $0$ otherwise. Thus, in one spatial dimension the roughness exponent ($3/2$) is much higher than that of the standard EW and KPZ classes ($1/2$).

\section{Growth by diffusing activators: modified KPZ equation}
\label{sec-11}

We now  study a stochastic lattice model with diffusing activators which  should display similar scaling to the field equations discussed in the previous section. There are, however, two crucial differences. 
First, the difference in height between neigbouring lattice sites is fixed to $1$ (solid-on-solid condition) the maximum possible width is $O(L)$, thus the maximum allowed value of $\alpha$ is $1$.
Second, when a continuum description is derived from the lattice model, non linear KPZ like terms may be generated. We shall discuss these issues below, but first we define the lattice model.

The model, introduced in~\cite{cagnetta2018aa}, consists of a discrete interface, made of $L$ unit-slope segments, and a collection of $N$ activators. Both the interface and the activators live on the one-dimensional lattice, with periodic boundary conditions enforcing the ring topology. Being made of unitary slopes, the interface can be described with a set of height variables $\left\lbrace h_i \right\rbrace$ over the lattice points $i=1,\dots,L$ which obey the solid-on-solid condition $|h_{i+1}-h_i|=1$. Each activator is represented by a discrete random walk hopping between neighbouring lattice sites. The activator dynamics is thus specified by the hopping rate $q$, which also coincides with twice the diffusion coefficient $2D_a$.

The interface, in turn, evolves according to a single-step dynamics. Due to the solid-on-solid condition, each site can be a peak ($\wedge$), a trough ($\vee$) or a slope ($\diagdown$ or $\diagup$). Troughs can grow and become peaks at rate $p_+$ whereas peaks become troughs at rate $p_-$, so that the solid-on-solid condition is preserved at all times. In order to account for the growth-stimulating action of the activators, we take the interface rates $p_{\pm}$ to depend on the local number of activators $n_i$, such that
\begin{equation}\label{eq:ModelIRates}
 p_-(i)=p,\quad p_+(i) = p + \lambda n_i.
\end{equation}
According to Eq.~(\ref{eq:ModelIRates}), each activator increases the growth rate of the interface by $\lambda$. As we assume no exclusion interaction among the activators, $n_i=0,\dots,N$.

\begin{figure}[h]
\includegraphics[width=\columnwidth]{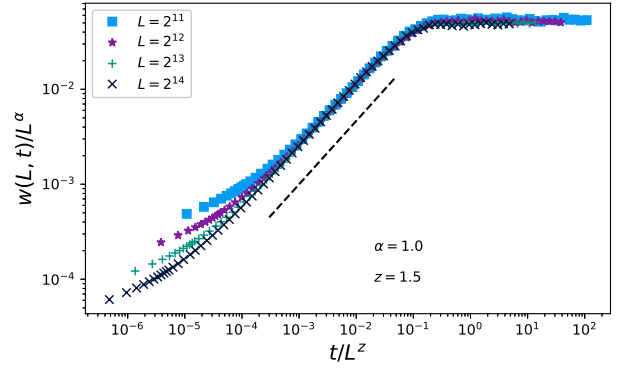}
\centering
\caption{Family-Vicsek scaling of the width $w(L,t)$ for the lattice model of active interface with diffusing activators, for $q=p=1$ and $\lambda=1$, system size $L$ as in the key. Averages are performed over $100$ independent realisations of the stochastic dynamics. The best collapse of the curves is achieved with $z=3/2$, as in the KPZ class, and $\alpha=1$, the maximal roughness of solid-on-solid models. The black dashed line is a guide to the eye for the $\sim t^{\alpha/z}$ law.}
\label{fig-1}
\end{figure}
In the lattice model, the width can be directly measured as the variance of the height variables $h_i$ over the lattice. We then track the time-dependent width of active interfaces of various sizes in Monte Carlo simulations of the update rule Eq.~(\ref{eq:ModelIRates}). The results are plotted in Fig.~\ref{fig-1} according to the Family-Vicsek scaling. Specifically, we found the best collapse to be achieved for $\alpha=1$ and $z=3/2$.

The scaling is different from that of the linear model discussed in the previous section. First the roughness exponent $\alpha$ takes its maximal value 1, which is less than the value $3/2$ for Eq.~(\ref{eq:activeedwardswilkinson}), though still higher than the EW- and KPZ-class exponent $1/2$.
Second, the value $3/2$ for the dynamic exponent suggests that a KPZ like nonlinearity is present. This can be traced back to the interface transitions occurring only at troughs and peaks in the solid-on-solid model, which implies a factor $1-(\nabla h)^2$ multiplying the driving force in Eq.~(\ref{eq:activeedwardswilkinson}):
\begin{equation}\begin{aligned}
\label{eq:activeKPZa}
\partial_t h(x,t) = &\lambda \left(1-(\nabla h)^2\right)\sum_{i=1}^N\delta\left(x-X_i(t)\right) +\\ &\nu \nabla^2 h(x,t) 
+ \sqrt{2D_h}\eta(x,t)
\end{aligned}\end{equation}
This term has the typical form (albeit with a difference in sign) of the normal projection factor of the KPZ equation, illustrated in Fig.~\ref{fig-2}, panel A. In fact, the added non-linear term can be shown to be relevant in the renormalisation group sense~\cite{cagnetta2020aa}, thus it is expected to change the scaling properties of the model.

\section{Slope-coupling due to normal growth}
\label{sec-2}

\begin{figure}[h]
\includegraphics[angle=-90,width=\columnwidth]{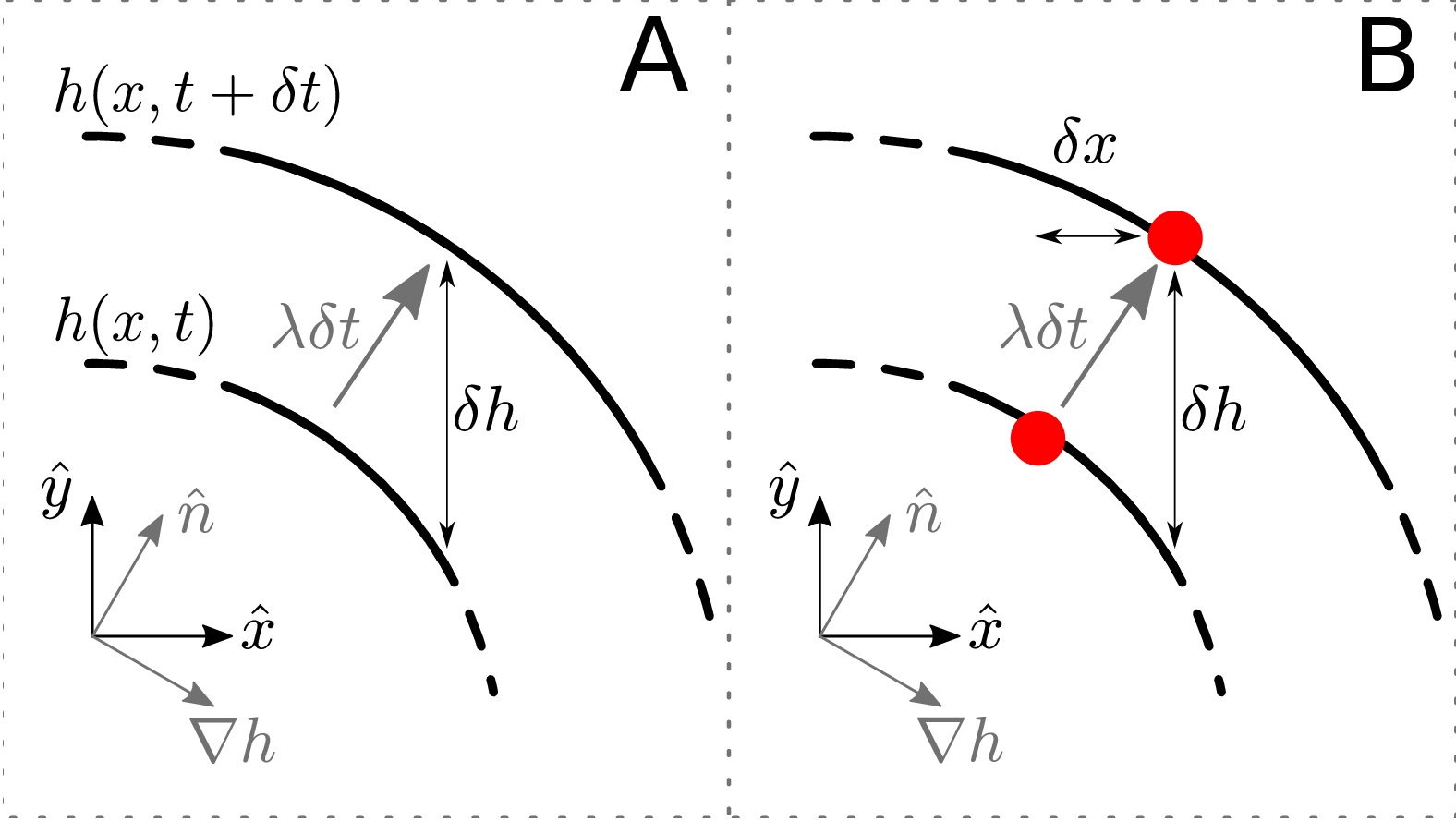}
\centering
\caption{Pictorial representation of the effects of normal growth on the active interface. Both pictures show the interface profile at two infinitely close instants, with the cartesian and local normal-tangent reference frames displayed in the bottom left corner, respectively in black and gray. Panel A: The displacement $\delta h$ due to a normal force $\lambda$ over time $\delta t$ reads $\delta h =\lambda\delta t [1+(\nabla h)^2]^{1/2}$. Panel B: The activator (red disk) is displaced by $\delta x = \lambda \delta t (\nabla h)[1+(\nabla h)^2]^{-1/2}$ due to the normal force acting on the interface.}
\label{fig-2}
\end{figure}

In this section we introduce a further modification of Eq.~(\ref{eq:activeedwardswilkinson}), which includes additional forces on the activators coming from geometric considerations~\cite{cai1995aa}.
An explanation for the origin of such terms is sketched in Fig.~\ref{fig-2}.
First, as in the KPZ equation~\cite{kardar1986aa}, local growth forces are exerted along the normal to the interface (see Fig.~\ref{fig-2}, panel A). The infinitesimal displacement $\lambda \delta t$ due to a force $\lambda$ over time $\delta t$ should then be increased by a factor $[1+(\nabla h)^2]^{1/2}\simeq 1+(\nabla h)^2/2$ (Fig.~\ref{fig-2}A). The same reasoning applies to a driving force which depends on the local density of activators along the interface, $\lambda \rho(x,t)$, where $\rho(x,t) =\sum_{i=1}^N\delta(x-X_i(t))$. According to the same argument, the matter which constitutes the interface is displaced horizontally by $\delta x = \lambda \delta t (\nabla h)[1+(\nabla h)^2]^{-1/2}$ (Fig.~\ref{fig-2}, panel B), generating an effective coupling of the activators positions with the interface slope.

We now write down equations for the density of activators $\rho(x,t)$ and an interface $h(x,t)$ driven by normal force $\lambda\rho(x,t)$. By keeping only terms which are at most quadratic in the fields, the resulting equations read
\begin{equation}\label{eq:activeKPZ}
\begin{aligned}
\partial_t h &= \lambda\rho +\frac{\lambda\rho_0}{2}(\nabla h)^2+ \nu \nabla^2 h + \sqrt{2D_h}\eta(x,t),\\
\partial_t \rho &= 2\lambda\rho_0 \nabla\left(\rho\nabla h\right) + D_a \nabla^2 \rho + \nabla\left(\sqrt{2D_a\rho_0}\xi(x,t)\right),
\end{aligned}
\end{equation}
with $\rho_0 = \int_{[0,L]^d}d^dx\,\rho(x,t)/L^d$ the global density of activators. We can represent the effective coupling of the activator positions with the interface slope via a minor modification of the lattice model. Instead of having the activators  hop left or right on the lattice at the same rate $q$, we let the rate depend on the difference in height between arrival and departure site, i.e.
\begin{equation}\label{eq:ModelARates}
 q\left(i \rightarrow i\pm 1\right) = \left\lbrace\begin{aligned} q_+,\quad&\text{  if } h_{i\pm 1} < h_i, \\q_-,\quad&\text{  if } h_{i\pm 1} > h_i. \end{aligned}\right.
\end{equation}
The parameter $\gamma=q_+-q_-$ controls the rate of slope advection, while $q=(q_++q_-)/2$ determines the mobility of the activators. The model defined by the rates of Eq.~(\ref{eq:ModelIRates}) and Eq.~(\ref{eq:ModelARates}) can be shown to be described by field equations analogous to Eq.~(\ref{eq:activeKPZ}), for the special choice of parameters $\lambda=2\gamma$~\cite{cagnetta2019ab}. Here we present the scaling of the interface width and the variance of the density of activators measured from Monte Carlo simulations of the lattice model.

\subsection{Width scaling}
\begin{figure}[h!]
  \begin{tabular}{c}
       \includegraphics[width=1\columnwidth]{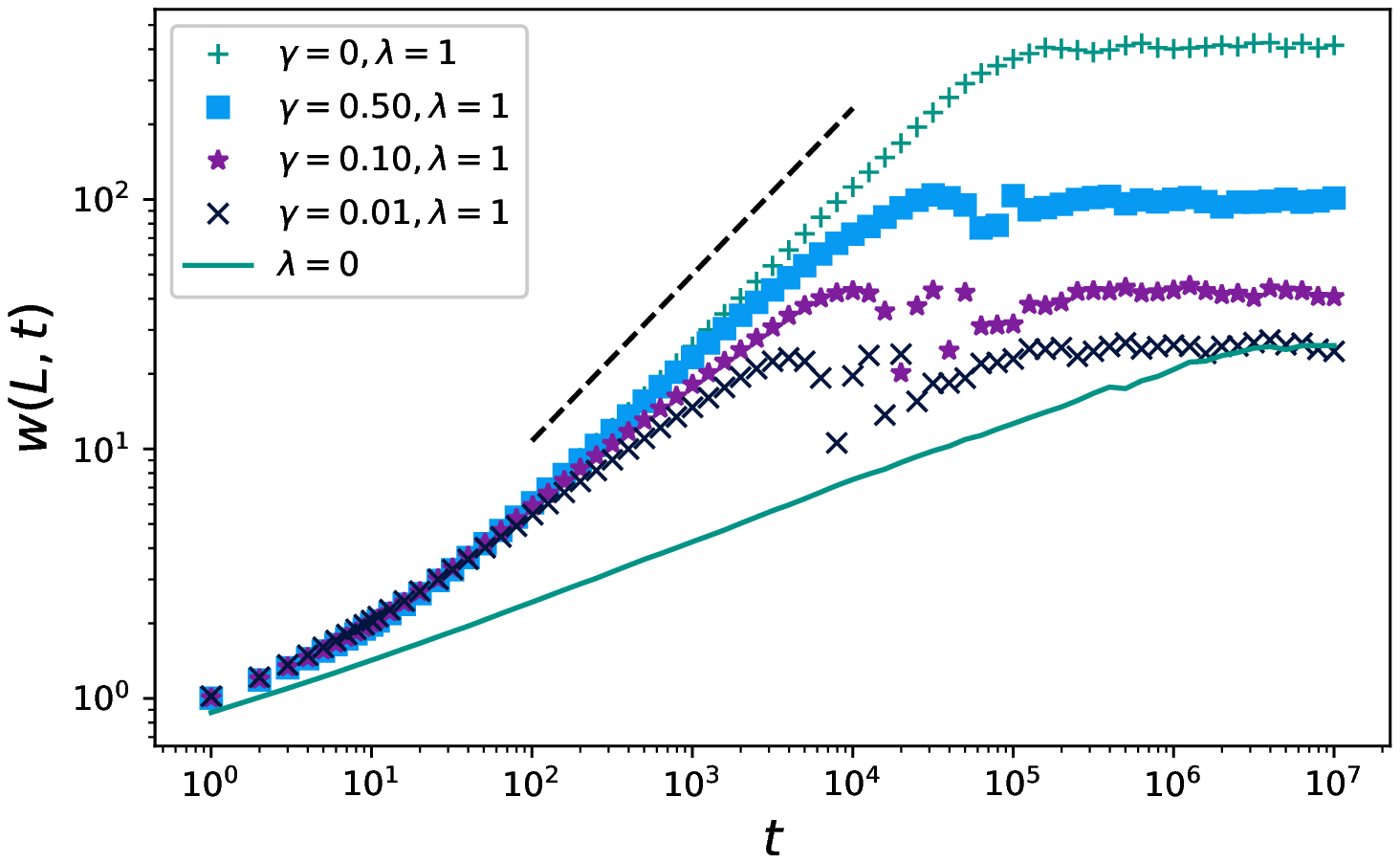}\\ \includegraphics[width=1\columnwidth]{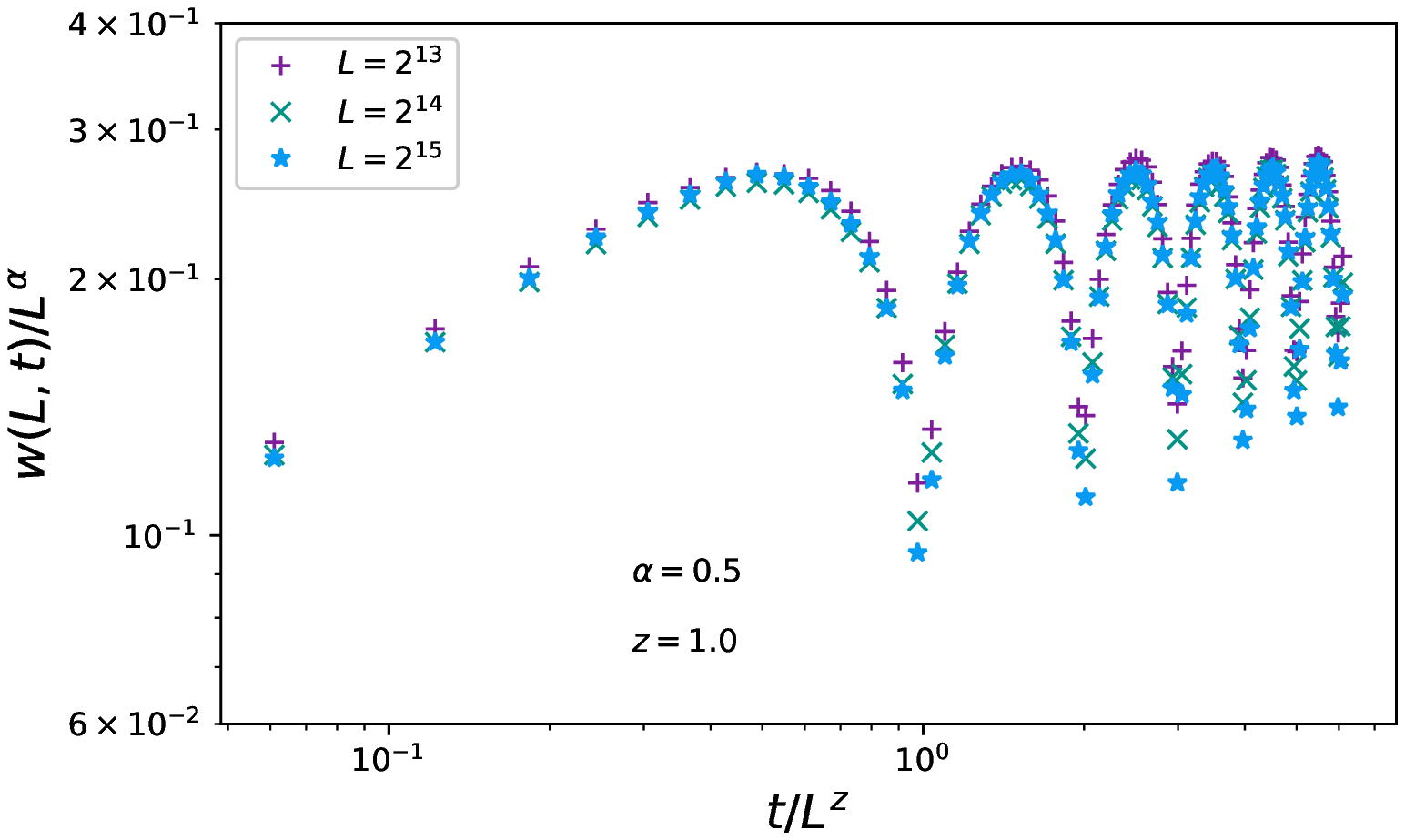}\\
  \end{tabular}  
\centering
\caption{Width of the active interface with slope-coupling $\gamma$. Top: comparison of the roughening profiles of interfaces with various $\gamma$, at $\lambda=1$, system size $L=2^{13}$ and density of activators $\rho_0=1$. The width of an EW interface is also shown for comparison as a teal solid line. The black dashed line is a guide to the eye representing the roughening law discussed in section \ref{sec-11}, $w\sim t^{2/3}$. Bottom: Family-Vicsek scaling of the active interface with $\gamma=0.5$, $\lambda=1$, density $\rho_0=1$ and $L$ as in the key. Here the oscillations are clearly visible. The best collapse is achieved for $\alpha=1/2$ and $z=1$. The exponents $z=1$ and $\alpha=1/2$ are typical of the whole $\lambda,\gamma > 0$ region of the parameter space.}
\label{fig-3}
\end{figure}

Let us begin with the width. The initial power-law increase, computed when starting from a flat initial condition, proceeds as in the case described in Section \ref{sec-11}, Fig.~\ref{fig-1}. Let us recall that the activators of the simulations shown in Fig.~\ref{fig-1} diffuse freely with no slope-coupling ($\gamma=0$). By turning on the slope-coupling, after some time which decreases for increasing $\gamma$, the width decreases and begins oscillating, until it reaches saturation. This property is manifest in Fig.~\ref{fig-3}, top panel, where the time-dependent widths of interfaces with different values of $\gamma\in[0,1]$ are compared. The width of an EW interface ($\gamma=\lambda=0$) is also shown for comparison (solid line in the figure).

The Family-Vicsek scaling (Eq.~(\ref{eq:familyvicsek})) is still obeyed for $\gamma\neq 0$, with a different set of exponents. As shown in Fig.~\ref{fig-3}, bottom panel, the roughness exponent $\alpha$ goes back to the EW and KPZ value $1/2$, typical of an interface with no correlations among the slopes in steady-state. However, the coupling of the activators position with the slopes results in an emergent ballistic behaviour, highlighted by the dynamic exponent $z=1$. This is a peculiar property of the active interface model introduced in~\cite{cagnetta2018aa} and can be explained by solving the inviscid limit of the field equations (\ref{eq:activeKPZ})~\cite{cagnetta2019ab}.

\subsection{Density scaling}
\begin{figure}[h!]
  \begin{tabular}{c}
       \includegraphics[width=1\columnwidth]{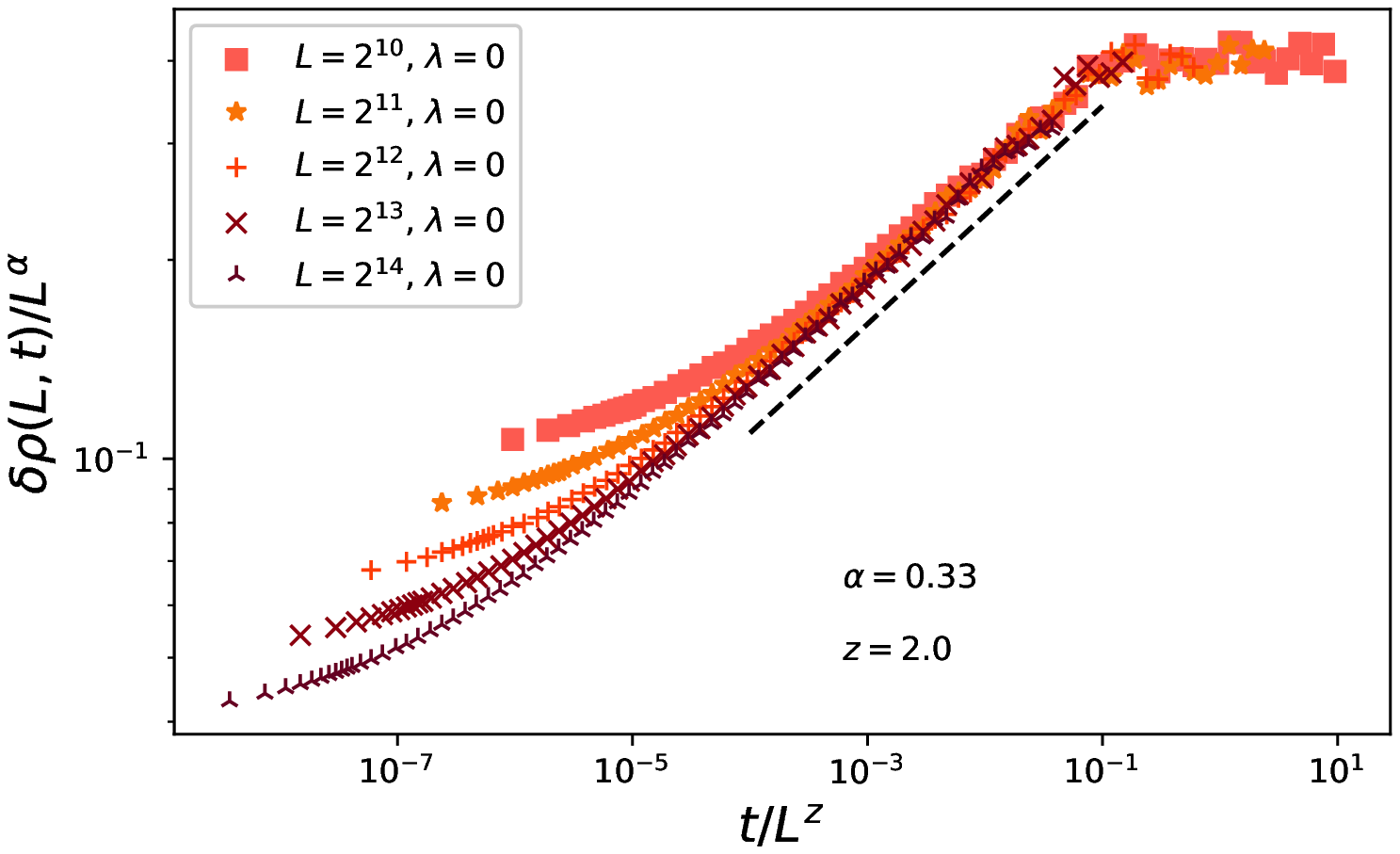}\\ \includegraphics[width=1\columnwidth]{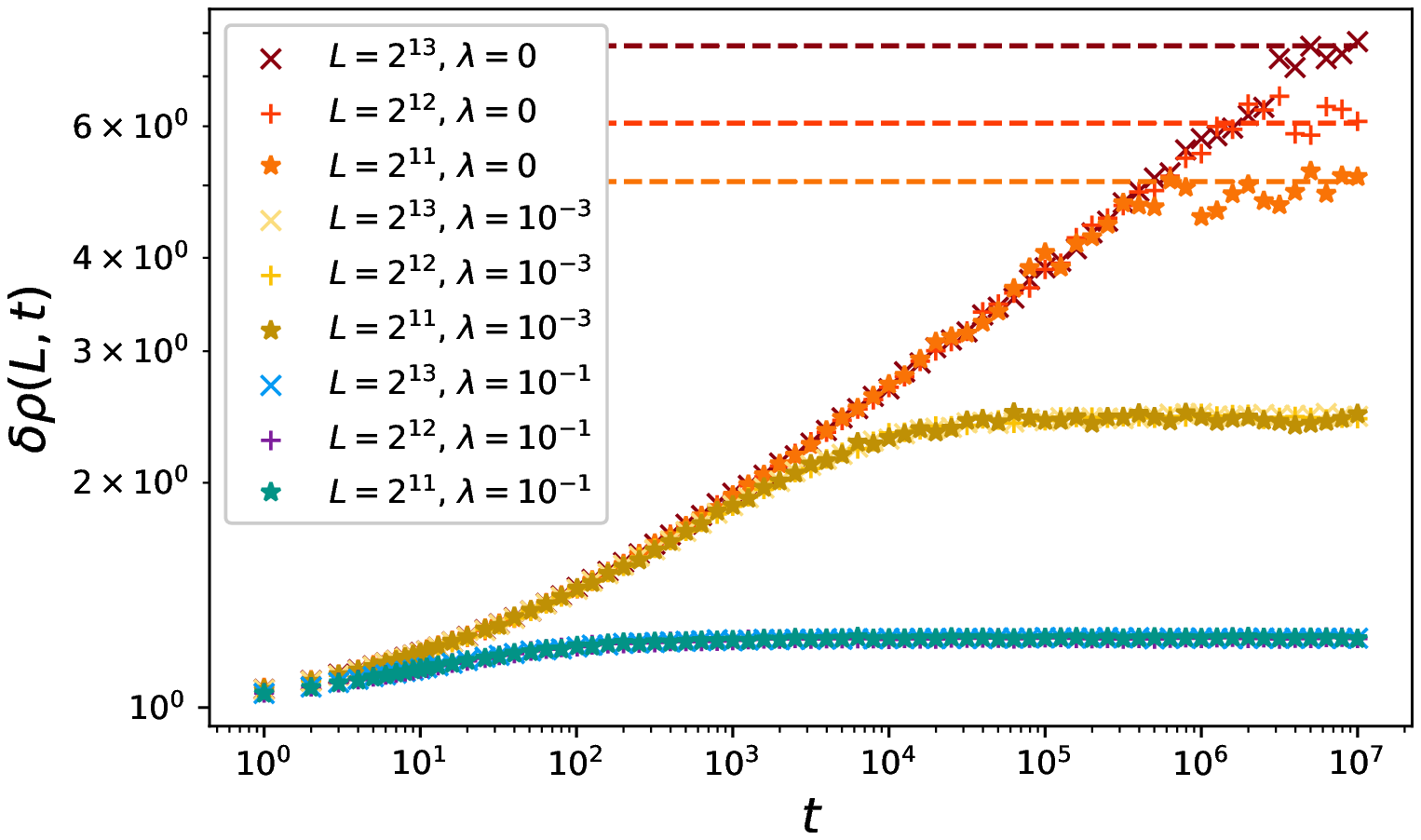}\\
  \end{tabular}  
\centering
\caption{Scaling of the density variance for a passive (top) and an active (bottom) interface. Parameters are $\gamma=0.5$, $\rho_0=1$, while $\lambda$ and $L$ are given in the key. In the passive case, the density variance obeys a scaling hypothesis analogous to the Family-Vicsek one, with dynamic exponent $z=2$ and ``roughness'' $\alpha=1/3$. For the density, the power-law initial growth in time and the power-law dependence on the system size of the steady-state value are evidences of coarsening in the system. In the active case, the scaling hypothesis is still obeyed albeit with trivial exponents $\alpha=z=0$.}
\label{fig-4}
\end{figure}

As the variance of the height profile provides a picture of the interface roughening dynamics, the variance of the activator density measures the activators coarsening. In the lattice model, where an occupation number $n_i(t)$ specifies the instantaneous number of activators at the $i$-th site, the density variance is defined as
\begin{equation}\label{eq:densvardef}
\delta\rho^2(L,t) = \frac{1}{L^d} \sum_{i=1}^L\avg{\left(n_i-\rho_0\right)^2}.
\end{equation}
Under appropriate circumstances, $\delta\rho$ obeys a scaling hypothesis analogous to that of Family-Vicsek for the width, Eq.~(\ref{eq:familyvicsek}), i.e. $\delta\rho(L,t)=L^{\chi}g(t/L^z)$. As in the Family-Vicsek scaling, the dynamic exponent describes the dependence of the density relaxation time on the size of the system. The exponent $\chi$, instead, by representing the scaling of the steady-state density variance with the systems size, indicates the extent of clustering in the system. If, for instance, macroscopic clustering takes place -- i.e., a finite number of sites hosts a finite fraction of the available particles -- then a few $n_i$'s scale as $N=\rho_0 L^d$ while all the others are close to zero, so that $\delta\rho^2$ will scale as $L^d$ and $\chi = d/2$. By contrast, for an homogeneous distribution of activators, $\delta\rho^2$ does not depend on the system size, i.e. $\chi=0$.

The scaling hypothesis of the density variance is obeyed in problems of non-interacting particles sliding down the slopes of a fluctuating interface~\cite{drossel2002aa,nagar2005aa,nagar2006aa,singha2018aa}. In the $\lambda=0$ limit of our lattice model, corresponding to passive particles sliding on a EW interface, we find $z=2$ and $\chi=1/3$ (cf. Fig.~\ref{fig-4}, top panel). When activity is turned on (bottom panel of Fig.~\ref{fig-4}), the density variance saturates at a finite value which is independent of the system size, indicating a homogeneous distribution of activators at large scales, i.e. $\chi=0$. The saturation time does not depend on the system size either, i.e. $z=0$. Nevertheless, the initial power-law growth of the density variance, especially visible at smaller values of $\lambda$, indicates that an initially homogeneous distribution of activators coarsens in time as in the passive sliders case. However, the size of the aggregates remains finite rather than growing with the system size---a phenomenon interpreted as \emph{microphase separation} in~\cite{cagnetta2018aa}.

\section{Conclusions}
\label{conclusion}
In this paper we have reviewed how interface growth equations and the resultant scaling are modified in the presence of diffusive particles which activate the growth.
We first studied analytically a modified EW equation with the addition of activators  and found novel scaling. However, this scaling is not observed in a simple simulation model  due to the presence of KPZ-like non-linearities. We then presented numerical results for the dynamics of activators coupled to the interface shape, which generalises the problem of passive scalar advection by fluctuating interfaces.
Our results indicate that a wealth of new dynamical behaviours are possible in the kinetics of active interfaces: scaling concepts might then be crucial in sorting them into different classes, as it is done for generic critical phenomena~\cite{HohenbergHalperinCriticalDynamics}.


\bibliography{proc}

\begin{thebibliography}{22}

\bibitem{halpin1995kinetic}
T.~Halpin-Healy, Y.C. Zhang, Physics reports \textbf{254}, 215 (1995)

\bibitem{barabasi1995}
A.L. Barabasi, H.E. Stanley, \emph{Fractal Concepts in Surface Growth}
  (Cambridge University Press, 1995)

\bibitem{eden1961growth}
M.~Eden, Dynamics of fractal surfaces \textbf{4}, 223 (1961)

\bibitem{turlier2016aa}
H.~Turlier, D.A. Fedosov, B.~Audoly, T.~Auth, N.S. Gov, C.~Sykes, J.F. Joanny,
  G.~Gompper, T.~Betz, Nature Physics \textbf{12}, 513 (2016)

\bibitem{brochard1975frequency}
F.~Brochard, J.~Lennon, Journal de Physique \textbf{36}, 1035 (1975)

\bibitem{prost1996active}
J.~Prost, R.~Bruinsma, EPL (Europhysics Letters) \textbf{33}, 321 (1996)

\bibitem{ramaswamy2000aa}
S.~Ramaswamy, J.~Toner, J.~Prost, Phys. Rev. Lett. \textbf{84}, 3494 (2000)

\bibitem{gov2006aa}
N.S. Gov, A.~Gopinathan, Biophysical Journal \textbf{90}, 454  (2006)

\bibitem{maitra2014aa}
A.~Maitra, P.~Srivastava, M.~Rao, S.~Ramaswamy, Phys. Rev. Lett. \textbf{112},
  258101 (2014)

\bibitem{cagnetta2018aa}
F.~Cagnetta, M.R. Evans, D.~Marenduzzo, Phys. Rev. Lett. \textbf{120}, 258001
  (2018)

\bibitem{cagnetta2019aa}
F.~Cagnetta, M.R. Evans, D.~Marenduzzo, Phys. Rev. E \textbf{99}, 042124 (2019)

\bibitem{family1985scaling}
F.~Family, T.~Vicsek, Journal of Physics A: Mathematical and General
  \textbf{18}, L75 (1985)

\bibitem{edwards1982surface}
S.F. Edwards, D.~Wilkinson et~al., Proc. R. Soc. Lond. A \textbf{381}, 17
  (1982)

\bibitem{kardar1986aa}
M.~Kardar, G.~Parisi, Y.C. Zhang, Physical Review Letters \textbf{56}, 889
  (1986)

\bibitem{cagnetta2020aa}
F.~Cagnetta, M.R. Evans, D.~Marenduzzo, V.~Skultety, \emph{Unpublished}  (2019)

\bibitem{cai1995aa}
W.~Cai, T.C. Lubensky, Phys. Rev. E \textbf{52}, 4251 (1995)

\bibitem{cagnetta2019ab}
F.~Cagnetta, M.R. Evans, Journal of Statistical Mechanics: Theory and
  Experiment \textbf{2019}, 113206 (2019)

\bibitem{drossel2002aa}
B.~Drossel, M.~Kardar, Phys. Rev. B \textbf{66}, 195414 (2002)

\bibitem{nagar2005aa}
A.~Nagar, M.~Barma, S.N. Majumdar, Phys. Rev. Lett. \textbf{94}, 240601 (2005)

\bibitem{nagar2006aa}
A.~Nagar, S.N. Majumdar, M.~Barma, Physical Review E \textbf{74}, 021124 (2006)

\bibitem{singha2018aa}
T.~Singha, M.~Barma, Physical Review E \textbf{98}, 052148 (2018)

\bibitem{HohenbergHalperinCriticalDynamics}
P.C. Hohenberg, B.I. Halperin, Rev. Mod. Phys. \textbf{49}, 435 (1977)

\end{thebibliography}

\end{document}